# Comment on "Radiative transfer in $CO_2$-rich atmospheres: 1. Collisional line mixing implies a colder early Mars"

M. Turbet and H. Tran

Laboratoire de Météorologie Dynamique, IPSL, UPMC Univ Paris 06, Ecole polytechnique, Ecole normale supérieure, Sorbonne Universités, Université Paris-Saclay, PSL Research University, CNRS, 4 place Jussieu, 75005, Paris, France

Corresponding author: Martin Turbet (martin.turbet@lmd.jussieu.fr)

**Abstract**

*Ozak et al. [2016]* claimed that explicitly including the effect of $CO_2$ collisional line mixing (LM) in their radiative transfer calculations yield $CO_2$ atmospheres that are more transparent to infrared radiation than when spectra calculations are made using sub-Lorentzian line shapes. This would in particular imply significantly colder surface temperatures (up to 15 K) for early Mars than estimated in previous studies. Here we show that the relative cooling that *Ozak et al. [2016]* associated to the effect of collisional line mixing is in fact due to a wrong choice of broadening species (air instead of $CO_2$). We then calculated Line-by-Line spectra of pure $CO_2$ atmospheres using a line-mixing model developed for self-broadened $CO_2$. Using the LMD Generic model (in 1D radiative-convective mode), we find that calculations made with the proper collisional line mixing model and with sub-Lorentzian line shapes lead to differences between early Mars surface temperatures smaller than 2 Kelvins only.

**Main text**

We first introduce here the various spectroscopic terms needed for the reader to understand the content of this comment. Modeling the entire $CO_2$ "allowed" spectrum (i.e. due to the molecule dipole and not to collision induced absorption), including both the regions near the lines centers and the far wings, is an extremely difficult task for which no rigorous model is available so far. The measured spectrum can be significantly different from that calculated with the usual Lorentz (or Voigt) profile, due to two effects that are neglected by this profile. The first one, called line mixing, is associated with the collisional transfers of rotational populations between absorption lines. It modifies the shape of clusters of closely spaced lines and results in transfers of absorption from the band wing region to the band center *[Hartmann et al. 2008]* leading to the strongly sub-Lorentzian behavior observed in $CO_2$ band wings. The second effect, related to the finite duration of collisions, affects the absorption in the far wings of the lines only *[Hartmann et al. 2008]*. To model $CO_2$ absorption spectral shape, two approaches are commonly used. In the first, called the χ-factor approach, an empirical correction of the Lorentzian shape is adjusted to laboratory measurements of the absorption in some band wings (e.g., at 4.3 μm in

*Perrin and Hartmann 1989*). The effects of both line-mixing and the finite collision duration are thus taken into account by this approach, but for the considered band wings region only. In the absence of precise and available data for other spectral region, the same χ-factor correction is generally used for all other $CO_2$ bands. In addition to this approximation, the effect of line mixing in the band center is not taken into account by the χ-factor correction. The second approach, based on the use of the impact and the Energy Corrected Sudden approximations (*Tran et al. 2011* and references therein), takes line mixing into account but not the effect of finite duration of collisions. This model, self-consistent for all bands, leads to satisfactory agreement with laboratory measurements for various bands, at different pressure and temperature conditions *[Tran et al. 2011]*. Very accurate predictions are obtained in the central regions of the bands with discrepancies that may increase together with the breakdown of the impact approximation *[Hartmann et al. 2008]*, when going far away in the wings. The line mixing and the χ-factor approaches are thus fully different and rely on completely different approximations. Yet, they can both be used to model broad band $CO_2$ absorption spectra.

*Ozak et al. [2016]* (hereafter, *OZ16*) explored the effect of using the line mixing (hereafter, LM) approach in $CO_2$-dominated atmospheres, typical of the early Martian environment (see review by *Forget et al. [2013]*). They found that, using a 1D radiative-convective model, the use of the LM approach results in colder early Mars surface temperatures than those obtained with the χ-factor approach. Note that, as mentioned above, LM effects on the line wings are already taken into account in the spectral calculations using the χ-factor approach, therefore, one cannot state that these calculations correspond to the case of "no line mixing", as done in *OZ16*. For instance, for a pure $CO_2$ atmosphere with a surface pressure of 2 bars, *OZ16* reported that surface temperatures could be lowered by about 15 K compared to previous studies *[Wordsworth et al. 2010]* in which the χ-factor approach was used. Such a result would have profound implications for the early Mars enigma *[Wordsworth et al. 2016, Haberle et al. 2017 – Chapter 17]*, making the formation of ancient valley networks and lakes even more difficult to explain.

However, here we show that this cooling is due to an improper choice of the LM model. *OZ16* used the updated version *[Lamouroux et al. 2010]* of the LM approach database and software package of *Niro et al. [2005]*, built for Earth atmosphere studies, i.e. for $CO_2$ broadened by air and not for pure $CO_2$, as needed for the case of early Mars. Here we instead use the LM package devoted to pure $CO_2$, developed by *Tran et al. [2011]* and updated in *Kassi et al. [2015]* to calculate absorption spectra of pure $CO_2$ under various pressure and temperature conditions typical of the early Martian environment.

An example of our calculated spectra is shown in Fig. 1 in which those obtained by *OZ16* are also reported for comparison. As in *OZ16*, we performed two calculations: the first with the LM approach but using the package of *Tran et al. [2011]* and *Kassi et al. [2015]* for pure $CO_2$ (hereafter denoted by pure $CO_2$ LM, in red). In the second calculation, the χ-factor approach for pure $CO_2$ *[Perrin and Hartmann 1989]* was used (in blue). Note that the latter is the widely-adopted procedure to calculate absorptions for early Martian $CO_2$-dominated atmospheres *[Wordsworth et al. 2010, Mischna et al. 2012]*. Finally, we also calculated spectra, as done by *OZ16*, with $CO_2$-in-air LM by using the package of *Niro et al. [2005]* and *Lamouroux et al. [2010]* (in green). Figure 1 calls for two remarks. Firstly, our calculations with $CO_2$-in-air LM and with the pure $CO_2$ χ-factor approach agree very well with those of *OZ16*. The second and very important remark is that when using the correct collision partner, i.e. pure $CO_2$ LM, the obtained spectrum (red) is closer to that calculated with the pure $CO_2$ χ factor approach (blue), with respect to the $CO_2$-in-air LM calculation (green).

Using the LMD Generic model (in 1D time-marching radiative-convective mode) parameterized following *OZ16*, with the two spectral calculation approaches for pure $CO_2$ (i.e., LM and χ factor), we performed numerical simulations of pure $CO_2$ atmospheres of various thicknesses (from 0.1 to 2 bars), under Noachian Mars conditions as in *OZ16*. Pure $CO_2$ continuum was taken into account in the two calculations. Specifically, the $CO_2$-$CO_2$ far infrared Collision Induced Absorptions (CIA) from *Gruszka and Borysow [1997]* and CIA and dimer absorptions from *Baranov et al. [2004]* were used and extrapolated for the whole spectral range, as widely done by previous studies and in particular by *Wordsworth et al. [2010]*. Fig. 2 shows the equilibrium surface temperatures obtained with (red) pure $CO_2$ LM and with (blue) the pure $CO_2$ χ-factor approach. For comparison, the results obtained by *OZ16* using $CO_2$-in-air LM (green) and those obtained in *Wordsworth et al. [2010]* (black) in which χ factors were used, were also plotted. In addition, we also performed simulations using $CO_2$-in-air LM. For this case, two calculations were made. In the first (solid orange line), pure $CO_2$ continuum was calculated as mentioned above, i.e. extrapolated for the whole spectral range. In the second, the same data were used but were truncated at 250 cm$^{-1}$ for far infrared CIA and between 1200 and 1600 cm$^{-1}$ for CIA and dimer absorptions from *Baranov et al. [2004]*, respectively, as done in *OZ16* (dashed orange line). As can be observed in Fig. 2, the result of this latter is in very good agreement with that of *OZ16*, indicating that the 1-D radiative-convective version of our LMD Generic Model agrees well with the one developed by *OZ16*. The correlated-k radiative transfer model of *OZ16* is thus not concerned by our comment and could be used for radiative transfer in planetary atmospheres.

Firstly, as can be observed, our simulations with the χ-factor approach (blue) agree well (±2K) with those of *Wordsworth et al. [2010]* confirming the calculation procedure we used. The remaining difference is likely due to the spectroscopic data used in the two calculations. Wordsworth et al. [2010] used data from HITRAN 2004 *[Rothman et al. 2005]* while in our calculations, HITRAN 2012 was used *[Rothman et al. 2013]*. Secondly, surface temperatures obtained with pure $CO_2$ LM and pure $CO_2$ χ factors are very close to each other, the differences being always smaller than 2 K for all considered surface pressures. This is consistent with the fact that absorptions calculated with pure $CO_2$ LM are very close to those calculated with the χ factors in the most relevant infrared spectral regions for early Mars, as shown in Fig. 1. This is also consistent with the fact that LM effects on the line wings are by nature already taken into account in the χ factor approach. In the opposite, high resolution spectra with $CO_2$-in-air LM (green, see Fig. 1) being more transparent in infrared regions, surface temperatures obtained in this case can be much lower (up to 10 K, solid orange line on Fig. 2). Note that in addition to the effect of the wrong broadening species, due to the effect of pure $CO_2$ continuum truncation, *OZ16* obtained even lower surface temperature (dashed orange and green lines on Fig. 2).

In summary, the significant cooling reported in *Ozak et al. [2016]* is mostly due to a wrong choice of broadening species (air instead of $CO_2$). Moreover, we show that early Mars surface temperatures calculated when using the proper LM model are very close to those obtained from spectra calculations based on the "usual" χ factor approach, their differences being smaller than 2 K, and thus within the uncertainty of usual early Mars radiative transfer calculations *[Ozak et al. 2016; Figure 2]*. This very good agreement thus justifies the use of the usual χ-factor approach for early Mars climate studies. This work also stresses the need for accurate spectroscopic data for early Mars pressure and temperature conditions, as well as for their careful use.

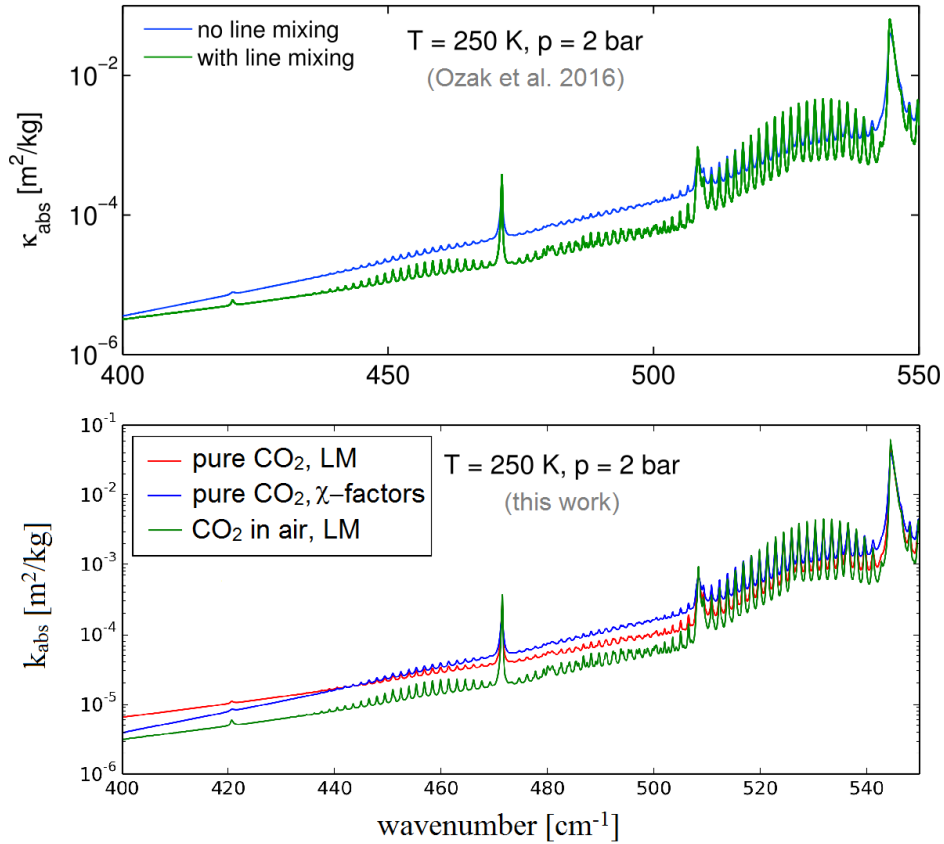

**Figure 1**. Line-by-line absorption spectra of a pure $CO_2$ atmosphere, at a temperature of 250K and a pressure of 2 bars. Upper panel: Figure directly imported from *OZ16*, calculated with their χ factors corrections (blue) or with their (green) inclusion of $CO_2$-in-air line mixing (LM). For comparison, we plotted in the lower panel 3 spectra calculated under the same pressure and temperature condition: in blue is the calculation with our χ factors, in green calculation with our $CO_2$-in-air line mixing, which both match the curves of *OZ16*, and in red our pure $CO_2$ line mixing. Absorption coefficient unit is in $m^2$ $kg^{-1}$, as in *OZ16*. Note that contribution of pure $CO_2$ continuum is not included here.

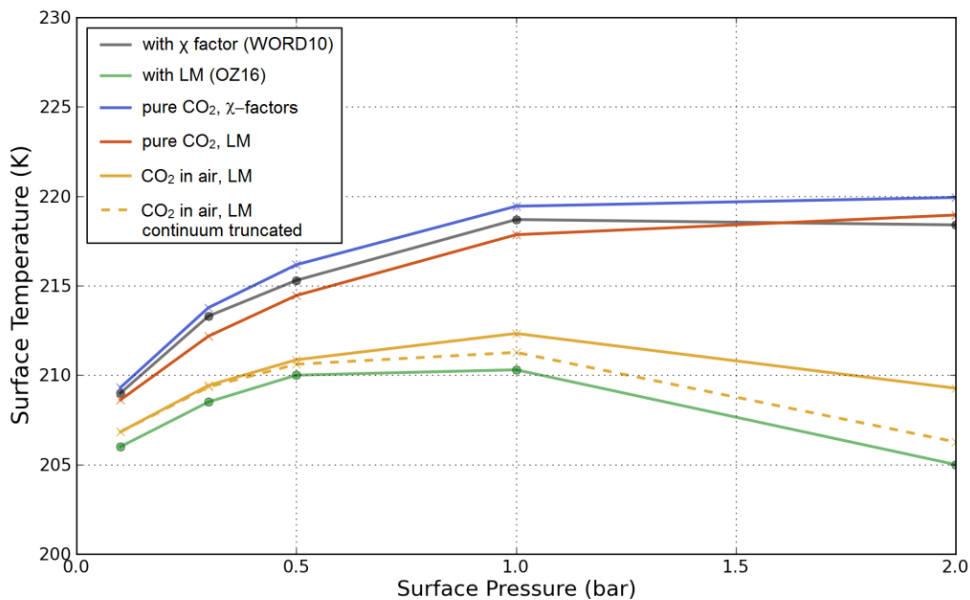

**Figure 2**. Temperatures at the surface of a pure $CO_2$ atmosphere of various thicknesses, exposed to early Mars conditions, and computed with the LMD Generic model, with pure $CO_2$ LM (red), and with the $\chi$ factors corrections (blue). Orange lines correspond to calculations using $CO_2$-in-air LM with pure $CO_2$ continuum truncated (dashed line) or not (solid line). For comparison, we plotted in black the original curve from *Wordsworth et al. [2010]* ($\chi$ factors), and in green the one from *OZ16* (with $CO_2$-in-air LM). Calculations include $CO_2$ condensation. Mean incident flux at the top of atmosphere is 442 W m$^{-2}$.

**Acknowledgments, Samples, and Data**

We thank François Forget and Jean-Michel Hartmann for fruitful discussions and remarks related to this work. We also acknowledge Itay Halevy, Oded Aharonson and Nataly Ozak for their constructive feedbacks. The code (LMD Generic Model) and correlated-k absorption coefficients developed in this study are both available upon request from the corresponding author.